\documentclass[a4paper,fleqn,usenatbib]{mnras}

\usepackage[T1]{fontenc}
\usepackage{ae,aecompl}

\usepackage{graphicx}	
\usepackage{rotating}
\usepackage{listings}
\usepackage{threeparttable}

\graphicspath{{./}{figs/}}

\usepackage{comment}
\usepackage{ulem,color}

\usepackage{natbib}

\usepackage{newtxtext,newtxmath}

\begin{document}

\title{Daily variability at milli-arcsecond scales in the radio quiet NLSy1 Mrk~110}

\author[Francesca Panessa]
{Francesca Panessa$^{1}$\thanks{Contact e-mail:\href{mailto:francesca.panessa@inaf.it}
{francesca.panessa@inaf.it}},
Miguel P\'erez-Torres$^{2,3}$,
Lorena Hern\'andez-Garc\'ia$^{4,5}$,
Piergiorgio Casella$^{6}$,
\newauthor
Marcello Giroletti$^{7}$,
Monica Orienti$^{7}$,
Ranieri D. Baldi$^{7,8}$,
Loredana Bassani$^{9}$,
Maria Teresa Fiocchi$^{1}$,
\newauthor
Fabio La Franca$^{10}$,
Angela Malizia$^{9}$,
Ian McHardy$^{8}$,
Fabrizio Nicastro$^{6}$,
Luigi Piro$^{1}$,
\newauthor
Federico Vincentelli$^{8,11}$,
David R.A. Williams$^{12}$,
Pietro Ubertini$^{1}$
\\
$^{1}$ INAF - Istituto di Astrofisica e Planetologia Spaziali, via Fosso del Cavaliere 100, I-00133 Roma, Italy\\
$^{2}$ Instituto de Astrof\'isica de Andaluc\'ia,Glorieta de la Astronom\'ia, s/n,E-18008 Granada, Spain\\
$^{3}$ Facultad de Ciencias, Universidad de Zaragoza,C/ Pedro Cerbuna 12,E-50009 Zaragoza, Spain\\
$^{4}$ Millennium Institute of Astrophysics, Nuncio Monse{\~{n}}or S{\'{o}}tero Sanz 100, Of. 104, Providencia, Santiago, Chile\\
$^{5}$ Instituto de F\'isica y Astronom\'ia, Universidad de Valpara\'iso, Av. Gran Breta\~na 1111, Playa Ancha, Chile\\
$^{6}$ INAF – Osservatorio Astronomico di Roma, Via Frascati 33, 00044, Monte Porzio Catone (Rome), Italy\\
$^{7}$ INAF - Istituto di Radioastronomia, via Gobetti 101, 40129 Bologna, Italy\\
$^{8}$ School of Physics and Astronomy, University of Southampton, Southampton, SO17 1BJ, UK\\
$^{9}$ INAF - Osservatorio di Astrofisica e Scienza dello Spazio Bologna, via Piero Gobetti, 93/3, 40129 Bologna, Italy\\
$^{10}$ Dipartimento di Matematica e Fisica, Universit\`a degli Studi Roma Tre, Via della Vasca Navale 84, I-00146, Roma, Italy\\
$^{11}$ Villanova University, Department of Physics, Villanova, PA 19085, USA\\
$^{12}$ Jodrell Bank Centre for Astrophysics, School of Physics \& Astronomy, 
\\
The University of Manchester,
Alan Turing Building, Oxford Road, Manchester M13 9PL, UK
}

\pubyear{2021}

\newcommand{\Nh}{N_{\rm H}}
\newcommand{\Nii}{[N {\sc ii}]}

\label{firstpage}
\pagerange{\pageref{firstpage}--\pageref{lastpage}}
\maketitle

\begin{abstract}
The origin of radio emission in the majority of Active Galactic Nuclei (AGN) is still poorly understood. Various competing mechanisms are likely involved in the production of radio emission and precise diagnostic tools are needed to disentangle them, of which variability is among the most powerful. For the first time, we show evidence for significant radio variability at 5 GHz at milli-arcsecond scales on days to weeks time scales in the highly accreting and extremely radio-quiet (RQ) Narrow Line Seyfert 1 (NLSy1) Mrk~110. The simultaneous {\it Swift}/XRT light curve indicates stronger soft than hard X-ray variability. The short-term radio variability suggests that the GHz emitting region has a size smaller than $\sim$180 Schwarzschild radii. The high brightness temperature and the radio and X-ray variability rule out a star-formation and a disc wind origin. Synchrotron emission from a low-power jet and/or an outflowing corona is then favoured.

\end{abstract}


\section{Introduction} \label{sec:intro}

The variety of accretion and ejection mechanisms in Radio-Quiet (RQ) Active Galactic Nuclei (AGN) can be effectively investigated via multi-frequency observations, especially by simultaneously monitoring of the radio and the X-ray emission. The X-ray radiation is produced in the vicinity of the super massive black hole (SMBH) from an interacting accretion disc and a hot corona system \citep{haardt}. On the other hand, the radio emission might be due to a combination of different processes: star-formation, low power jet, shocks in outflows and/or coronal emission (see \citet{panessa19} for a review). Sub-pc radio emission in RQ AGNs may indeed be produced by a scaled-down version of more powerful jets, perhaps less collimated and with a lower acceleration efficiency. The finding of significant correlation between the radio and X-ray emission in both super-massive and Galactic black holes has suggested a strong coupling between the inflowing accretion and the outflowing ejection components \citep{panessa07,laor08}. For instance, it has been proposed that the X-ray corona itself could coincide with the jet base \citep{markoff05}.
Discriminating among the above different mechanisms is crucial to reach the heart of the matter of the disc-jet coupling.

The ideal experiment would combine a well time sampled monitoring with very high spatial resolution radio observations at high frequency together with simultaneous X-ray observations. Indeed, Very Long Baseline Interferometry (VLBI) studies are able to map radio emission at milli-arcsecond (mas) scales, sampling down to sub-pc scales regions in local Seyfert galaxies, resolving out diffuse emission and disentangling thermal versus non thermal emission \citep{nagar02,giroletti09,baldi18lem,baldi21}. Small structures close to the AGN core can also be mapped via high frequency observations \citep{Behar15}. 

In this framework variability studies are a powerful diagnostic tool for the accretion/ejection physics, but largely unemployed so far in RQ AGN. 
Radio variability has been tested only in a sparse number of RQ AGN \citep{wrobel00, barvainis05, anderson05, mundell09, bell11, doi11, king13, baldi15, williams20, behar20} and it was found to be typically of a few tens per cent over months to years time scales. 
In a few Low Luminosity AGN (LLAGN), using Very Large Array (VLA) observations, variability was found on timescales from a few hours to 10 days \citep{anderson05}. Variability on a time scale of the day has also been detected in the Seyfert 1 NGC\,7469 at 95 and 143 GHz; the same variability pattern, observed both in the mm and in the X-rays bands supported the idea that both emissions are originating in the same physical component of the AGN, likely the accretion disc corona \citep{baldi15, behar20}.

Here we report on the simultaneous radio and X-ray monitoring of Mrk~110 at 4.9 GHz with the Very Long Baseline Array (VLBA) and {\it Swift}/XRT. Mrk~110 (z=0.0353) is a RQ Narrow Line Seyfert 1 galaxy (NLSy1) which shows extreme variability at X-rays as well as at optical/UV wavelengths \citep{grupe01, kollatschny01}. It is one of the most variable AGN among the AGN RXTE sample on years to months time scales \citep{markowitz04}. Mrk~110 is also variable on smaller time scales in X-rays, as revealed during a 47.5 ks {\it XMM-Newton} observation in 2004, where the size of X-ray emitting region was constrained to be R $\sim$ 10 R$_{s}$ -- Schwarzschild radii -- \citep{dasgupta06}.
At radio mas scales, \citet{doi13} have revealed the presence of a compact core at 1.7 GHz with a flux density of 1.2$\pm$0.2 mJy. The high brightness temperature (T$_{B}$ $>$ 6.5 $\times$ 10$^{7}$) suggested non-thermal emission possibly coming from a jet-base.

\section{Observations and results}

\begin{figure*}
\centering
{\includegraphics[width=0.83\textwidth]{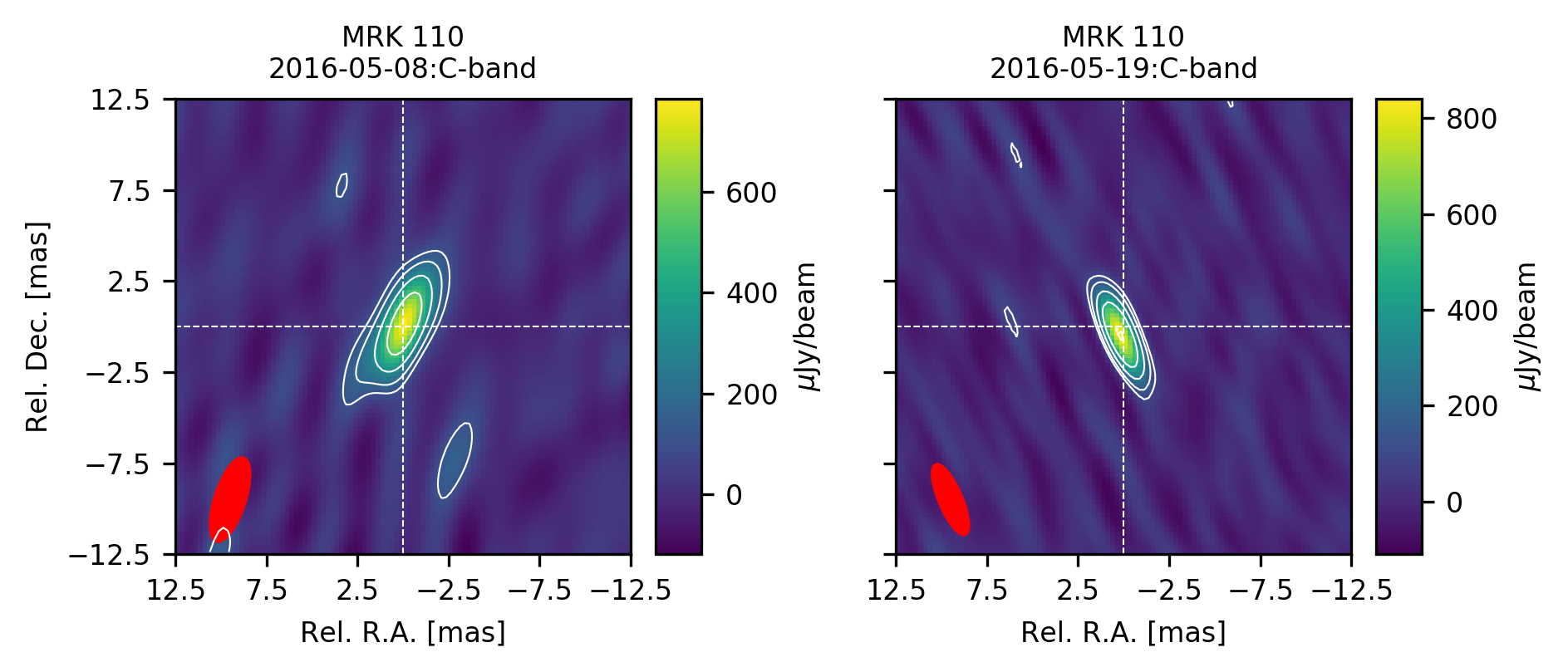} }
\caption{VLBA images corresponding to the epochs BP196N and BP203BA. The images are centered at RA(J2000.0) $=$ 09h25m12.8477400s and Dec(J2000.0) $=$ +52$^\circ$17$^\prime$10.387571$^{\prime\prime}$. Contours are drawn at (3, 3$\times$ $\sqrt{3}$, 9, 9$\times$ $\sqrt{3}$,..., 27) $\times$ rms (which is 36 and 30 $\mu$Jy beam$^{-1}$ respectively, as reported in Table ~\ref{tab:radio}). These two epochs were selected to show an example of a marginally resolved (left) and of an unresolved (right) core emission.}
\label{fig:radio}
\end{figure*}

\subsection{Radio observations}

We present VLBA observations taken during a monitoring campaign of Mrk 110 from 2015 August 14 till 2016 May 19. A nearly daily cadence monitoring was performed at 4.9 GHz between 2016 April 24 and May 19 (project code BP196). Contemporaneous VLBA 
observations on 2015 August 14 (project code BP193) at 1.6 (L-band), 4.7 and 7.4 GHz were performed (C- and X-band respectively). VLBA data at 1.6 and 5 GHz taken in April 2016 (project code BP203), during two consecutive days, are also considered. The log of the VLBA observations are shown in Table~\ref{tab:radio}. Uncertainties, unless otherwise stated, are at 68\% (1 sigma) confidence level.

Given the faintness of Mrk~110 at GHz frequencies, we referenced our VLBA observations with the nearby (1.4 deg separation) VLBA calibrator, J0932+5306. 
We used the NRAO AIPS package for all calibration and hybrid mapping and applied standard data reduction procedures to image Mrk~110.
In particular, we used the AIPS task \texttt{jmfit} to extract the peak and total flux densities by fitting the emission from Mrk~110 to a two-dimensional Gaussian. In our variability analysis, we conservatively took as total uncertainties the larger value of the following two estimates: the uncertainty given by task \texttt{jmfit}, or the sum in quadrature of the off-source root-mean-square (rms) of the VLBA flux density values and the calibration uncertainty, $\xi$, assumed to be of 5\% of the flux density $S$, i.e., $\sigma = \sqrt{{\rm rms}^2 + (\xi\,S)^2}$.

We detect an unresolved core from the maps of all VLBA epochs, with the possible exception of the epoch 8th May, when an elongated structure in the south-west direction is marginally resolved (see Fig.~\ref{fig:radio}). The peak flux densities range between 0.670 and 0.990 mJy/beam at 5 GHz and between 0.857 and 1.459 mJy/beam at 1.5 GHz, corresponding to powers of 2.0 -- 2.9 $\times$ 10$^{21}$ W/Hz and 2.5 -- 4.3 $\times$ 10$^{21}$ W/Hz respectively, for an assumed luminosity distance of 156.2 Mpc to Mrk~110. In one case, BP193 at X-band, the source was heavily unresolved, which resulted in the peak flux density being higher than the total flux density.

The flux density of our VLBA phase-calibrator, J0932+5306, stayed 
constant at 200$\pm$12 mJy for all epochs, but for the last two ones, where it dropped slightly, to a flux density of 175$-$180 mJy. This flux density light curve does not significantly correlate with the Mrk~110 total and peak flux densities (correlation coefficients of 0.60 and 0.21, respectively), hence excluding that the observed variability is inherited from the phase-calibrator behaviour.

The variability of the innermost regions (the core) of Mrk 110, probed by its 5 GHz peak flux density, exhibits variation amplitudes of 10\% throughout our campaign (Fig.~\ref{fig:flux}). The offset between the radio total and peak flux densities follow the same trend  indicating that the total flux density variability is mostly due to variability at very compact scales, except for the 8th May epoch, when there is marginal evidence for variability outside the innermost compact, unresolved region of radio emission. We therefore checked carefully the VLBA data for this epoch, which were of similar quality as the data obtained for the rest of the campaign. One-component Gaussian fit suggests the presence of a resolved component, indeed the use of two Gaussian components improves the fit significance, although only marginally. 

At the distance of Mrk~110, 1 mas corresponds to $\sim$ 0.71 pc. 
Therefore, the most compact measured beam sizes of 12.53 mas $\times$ 5.04 mas  (L-band), 3.81 mas $\times$ 1.36 mas (C-band) and 2.57 mas $\times$ 0.84 mas (X-band) correspond to linear sizes of 8.6 pc $\times$ 3.6 pc, 2.7 pc $\times$ 1.0 pc, and 1.8 pc $\times$ 0.6 pc, at L-, C- and X-band, respectively. The derived lower limits to the brightness temperatures range between 7.5 and 15.7 $\times$ 10$^{7}$ K. 

From the peak flux density values for Mrk~110 in col. 8 of Table~\ref{tab:radio}, we estimated two-point spectral indices, S$_\nu \sim \nu^{-\alpha}$. We estimated the uncertainty in the spectral index, $\sigma_\alpha$, by using the error propagation equation, which yields  $\sigma_\alpha = \sqrt{{{(\sigma_{\nu_1}/S_{\nu_1})}^2}+{{(\sigma_{\nu_2}/S_{\nu_2})^2}}}/ln({\nu_2/\nu_1})$, where $S_{\nu_{1,2}}$ and $\sigma_{\nu_{1,2}}$ are the flux density and its uncertainty at each frequency.
The variability of the day-to-day radio spectral index alpha, computed at 1.55 and 4.98 GHz, results to be not significant ($\alpha = -0.48\pm0.08$ and $-0.43\pm0.07$ on 18 and 19 May 2016, respectively). On the other hand, the two-point spectral index on 14 Aug 2015 between 1.55 and 4.40 GHz is $-0.15\pm 0.11$, a much flatter value. On the same date, the spectral index between 1.55 and 7.60 GHz is $-0.39\pm0.10$. Finally, the rather steep value of $\alpha = -0.85\pm 0.29$ between 4.40 and 7.60 GHz suggests important synchrotron losses of the electrons at frequencies above 4.4 GHz. 
\\

\begin{table*}
\begin{center}
\caption{\bf Log of the VLBA observations}
\begin{tabular}{lllrrrrrrrrr}
\hline \hline

\hline \hline 
Date & Code  &$\nu$  & bmaj &  bmin &  bpa &  S$_{\rm peak}$&    S$_{\rm peak beam}$  & S$_{\rm total}$  & S$_{\rm cal}$  & rms   \\
 1 & 2 &3 & 4 & 5 & 6 & 7 & 8 & 9 & 10 & 11 \\
           \hline

 2015-08-14 &   BP193\_L     &  1.55 &	 13.07 &  4.14 &  10.8  &857 $\pm$	65&	  & 	   1305 $\pm$	130&  & 49		    \\
 2015-08-14 &   BP193\_C    &  4.40 &	   4.61 &  1.48 &  -2.8 & 735 $\pm$	62&	  & 	    819 $\pm$	102&   &50		 \\
 2015-08-14 &   BP193\_X    &  7.60 &	   2.57 &  0.85 &  -2.2 & 463 $\pm$	65&	  & 	    398 $\pm$	97  &   &61		\\
 2016-04-24 &   BP196A\_C &  4.98 &	   4.57 &  1.14 & -13.8  &825 $\pm$	48& 1020 $\pm$ 56	  &  	   1029 $\pm$	57&     199.2 $\pm$  10.0  &   24    \\
 2016-04-25 &   BP196B\_C &  4.98 &	   4.50 &  1.32 & -19.4  &955 $\pm$	57& 1022 $\pm$ 60    &  	    986 $\pm$	58&     212.1 $\pm$  10.6  &   31     \\
2016-04-27 &   BP196C\_C &  4.98 &	   4.14 &  1.25 & -18.7 &990 $\pm$	65& 1237  $\pm$ 75   &  	   1118 $\pm$	70&     207.5 $\pm$  10.4  &   42 		\\
 2016-04-28 &   BP196D\_C &  4.98 &          4.10 &  1.16 & -12.5 &785 $\pm$	51&	890  $\pm$  55   &  	    838 $\pm$	53&     189.0 $\pm$  9.5    &   33 		 \\
 2016-04-29 &   BP196E\_C &  4.98 &	    4.18 &  1.21 & -12.8 &768 $\pm$	49&	918  $\pm$  55   &  	    918 $\pm$	55&     202.3 $\pm$  10.1  &   31 		\\
 2016-04-30 &   BP196F\_C &  4.98 &	    4.14 &  1.20 & -12.9  &836 $\pm$	51&	991  $\pm$  58   &  	    876 $\pm$	53&     199.8 $\pm$   10.0  &   30  \\
 2016-05-01 &   BP196G\_C &  4.98 &	    5.40 &  1.20 & -12.7 &838 $\pm$	52&	992  $\pm$  58  &  	    928 $\pm$	55&     213.1 $\pm$  10.7  &   30  \\
 2016-05-02 &   BP196H\_C &  4.98 &	    4.14 &  1.21 & -12.6 &838 $\pm$	52&	994  $\pm$  58  &  	    956 $\pm$	56&     205.3 $\pm$  10.3  &   30  \\
 2016-05-03 &   BP196I\_C  &  4.98 &	    4.93 &  1.17 & -15.4 &724 $\pm$	50&	868  $\pm$  56  &  	    861 $\pm$	55&     205.7 $\pm$  10.3  &   35  \\
 2016-05-04 &   BP196J\_C &  4.98 &	    4.97 &  1.08 & -14.7 &670 $\pm$	49&	699  $\pm$ 50   &  	    716 $\pm$	51&     189.1 $\pm$   9.5   &   36  \\
 2016-05-05 &   BP196L\_C &  4.98 &	    4.79 &  1.12 & -13.3 &671 $\pm$	48&	767  $\pm$ 52   &  	    705 $\pm$	50&     191.4$\pm$   9.6    &   35  \\
 2016-05-07 &   BP196M\_C &  4.98 &	   5.14 &  1.09 & -11.6 &708 $\pm$	50&	799  $\pm$ 54    &  	    789 $\pm$	53&     194.9 $\pm$   9.7   &   36    \\
 2016-05-08 &   BP196N\_C &  4.98 &	    4.79 &  1.65 & -18.1 &742 $\pm$	52&	945  $\pm$ 59    &  	   1080 $\pm$	65&     214.3 $\pm$  10.7  &   36     \\
 2016-05-18 &  BP203AA\_C &  4.98 &	   3.82 &  1.37 &  25.4 &836 $\pm$	56&	801  $\pm$ 55    &  	    848 $\pm$	56&     178.7 $\pm$   8.9   &   37     \\
 2016-05-18 &  BP203AB\_L &  1.55 &	  12.53 &  5.04 &  13.1 &1459 $\pm$	86&	  & 	   1668 $\pm$	120&  & 44		\\  				
 2016-05-19 &  BP203BA\_C &  4.98 &	   4.14 &  1.24 &  23.1 &864 $\pm$	53&	865  $\pm$ 53    &  	    873 $\pm$	54&     173.5 $\pm$ 8.7     &   30  \\ 
 2016-05-19 &  BP203BB\_L &  1.55 &	  12.71 &  5.04 &  13.3 &1426 $\pm$	83&	  & 	   1629 $\pm$	114& &  42				 \\

\hline
\end{tabular}
\label{tab:radio}
\end{center}
Table notes: (1) Observing date; (2) VLBA project code; (3) Frequency (GHz); (4-5) Beam major (and minor) axis, in mas; (6) Beam position angle, in degrees; (7) Peak, (8) Peak common beam (reported only for the data set used for the variability analysis, see Section 3) and (9) Total flux density in $\mu$Jy, uncertainties are calculated as the sum in quadrature of the off-source rms of the VLBA flux density values and the calibration uncertainty, $\xi$, assumed to be of 5\% of the flux density $S$, i.e., $\sigma = \sqrt{{\rm rms}^2 + (\xi\,S)^2}$; (10) Peak flux density of the phase-calibrator in the C-band epochs used for the variability analysis, in $\mu$Jy beam$^{-1}$; (11) off-source rms values in $\mu$Jy beam$^{-1}$.
 \end{table*}

\subsection{X-ray observations}

During the VLBA observations, the source was followed-up in quasi-simultaneity with {\it Swift}/XRT for 21 epochs. The data reduction of the {\it Swift} X-ray Telescope \citep[XRT,][]{burrows2005} was performed by following standard routines described by the UK {\it Swift} Science Data Centre (UKSSDC)\footnote{http://www.Swift.ac.uk/analysis/xrt/index.php} and using the software in HEASoft version 6.26. Calibrated event files were produced using the routine {\sc xrtpipeline}, accounting for bad pixels, vignetting effects and exposure maps. Spectra were extracted from circular regions of diameter of 40$\arcsec$, centered in the source position given by NED using the {\sc xselect} tool within FTOOLS. Background spectra were extracted from a number of empty circular regions, having 80" of diameter, located nearby each of the sources. The {\sc xrtmkarf} task was used to create the corresponding Ancillary Response Files (ARF). The response matrix files (RMF) were obtained from the HEASARC CALibration DataBase (CALDB). 

To systematically search for X-ray variations we applied the method described in \cite{lore2013, lore2015}, we refer the reader to these papers for details on the procedure. We fitted all the X-ray spectra of Mrk\,110 simultaneously using a simple model consisting on two absorbed power-laws plus galactic absorption (N$_{Gal}=1.47 \times 10^{20} $cm$^{-2}$, \citealt{dickey1990}) in XSPEC v.12.10.1 as wabs*(zwabs*po+zwabs*po). We note that leaving the intrinsic column density free to vary does not provide an improvement in the significance of the fit. The model which best represents the data was obtained when the normalizations of the two power-laws are linked and vary together (the intrinsic column density is fixed at a value of 8.5$^{+1.8}_{-1.7}$ $\times$ 10$^{21}$ cm$^{-2}$). The soft (0.5--2 keV) and hard (2--10 keV) unabsorbed fluxes resulting from the best fit can be found in Table~\ref{tab:xray} and are plotted in Fig.~\ref{fig:flux}.

\begin{table}
\begin{center}
\caption{\bf Log of the {\it Swift}/XRT observations quasi-simultaneous to VLBA}
\begin{tabular}{llll}
\hline \hline 
  Date    & ObsID & \multicolumn{1}{c}{F$_{\rm 0.5-2 keV}$} & \multicolumn{1}{c}{F$_{\rm 2-10 keV}$} \\
     1&2&3&4\\
     \hline
     
 2016-04-24 & 00092396001 &  1.80 $_{-0.05}^{+0.06}$ & 3.6 $_{-0.2}^{+0.2}$   \\
 2016-04-25 &  00092396002 &  1.89 $_{-0.06}^{+0.05}$ & 3.2 $_{-0.2}^{+0.2}$   \\
 2016-04-26 & 00092396003 &  1.63 $_{-0.05}^{+0.05}$ & 3.0 $_{-0.2}^{+0.2}$	\\  
 2016-04-27 & 00092396004 &   1.66 $_{-0.04}^{+0.05}$ & 3.5 $_{-0.2}^{+0.2}$   \\
 2016-04-28 & 00092396005 &   1.53 $_{-0.04}^{+0.05}$ & 3.2 $_{-0.2}^{+0.2}$   \\
 2016-04-29 & 00092396006 &   1.52 $_{-0.04}^{+0.05}$ & 3.1 $_{-0.1}^{+0.1}$   \\
 2016-04-30 & 00092396007 &   1.49 $_{-0.04}^{+0.05}$ & 2.8 $_{-0.2}^{+0.2}$   \\
 2016-05-01 & 00092396008 &   1.72 $_{-0.06}^{+0.05}$ & 3.2 $_{-0.2}^{+0.2}$   \\
 2016-05-02 & 00092396009 &   1.73 $_{-0.05}^{+0.04}$ & 3.3 $_{-0.2}^{+0.2}$   \\
 2016-05-03 & 00092396010 &   1.67 $_{-0.04}^{+0.05}$ & 3.2 $_{-0.2}^{+0.2}$   \\
 2016-05-04 & 00092396011 &   1.64 $_{-0.05}^{+0.05}$ & 3.2 $_{-0.2}^{+0.2}$   \\
 2016-05-05 & 00092396012 &   1.81 $_{-0.05}^{+0.07}$ & 3.7 $_{-0.3}^{+0.3}$   \\
2016-05-06  & 00092396013 &   1.58 $_{-0.04}^{+0.05}$ & 3.2 $_{-0.2}^{+0.2}$   \\  
 2016-05-07 & 00092396014 &   1.53 $_{-0.04}^{+0.05}$ & 3.3 $_{-0.2}^{+0.2}$ \\
 2016-05-08 & 00092396015 &   1.57 $_{-0.06}^{+0.04}$ & 3.4 $_{-0.2}^{+0.2}$ \\
2016-05-09  & 00092396016 &   1.58 $_{-0.06}^{+0.06}$ & 3.0 $_{-0.2}^{+0.2}$   \\ 
2016-05-17  & 00037561005 &  2.46 $_{-0.06}^{+0.06}$ & 3.5 $_{-0.2}^{+0.2}$   \\ 
 2016-05-18 & 00037561006 &  2.35 $_{-0.06}^{+0.06}$ & 3.3 $_{-0.2}^{+0.2}$   \\
 2016-05-19 & 00037561007 & 2.82 $_{-0.07}^{+0.07}$ & 3.7 $_{-0.2}^{+0.2}$   \\ 
2016-05-20  & 00037561008 &   2.52 $_{-0.07}^{+0.06}$ & 4.2 $_{-0.2}^{+0.2}$   \\
2016-05-21  & 00037561009 &   2.03 $_{-0.08}^{+0.09}$ & 3.1 $_{-0.3}^{+0.3}$   \\ 
\hline
\end{tabular}
\label{tab:xray}
\end{center}
(1) Observation Date; (2) Swift observation ID; (3) Soft 0.5--2 keV and (4) hard 2--10 keV unabsorbed X-ray fluxes in 10$^{-11}$ ergs/s/cm$^{2}$ (errors are given at 90\% confidence level).
 \end{table}

\section{Variability Analysis}
 
We show the results of our variability analysis in Table~\ref{tab:variability} for our radio and X-ray light curves obtained in April-May 2016. We list the mean flux density and its standard deviation. The $\chi^2$ of the light curve with respect to a constant flux is also presented with the degrees of freedom (d.o.f). In all cases the reduced-$\chi ^2$ ($\chi ^2_r$=$\chi ^2$/d.o.f) is $>>$ 1, suggesting the presence of intrinsic variability.
We follow \citet[]{sanchezsaez2017}, which define an expression of the variability probability, (P$_{var}$) = 1 - P($>\chi ^2$)), where P($>\chi ^2$) is the probability to observe a $\chi ^2$ larger than measured, under the null hypothesis of random errors and no variability. P($>\chi ^2$) lower than 0.05 (i.e., P$_{var}$ $>$ 0.95) is usually considered as a significant indication of the presence of variability. 

In addition, we have estimated the normalized excess variance, $\sigma^{2}_{NXS}$ and its error err($\sigma^{2}_{NXS}$), which represents the variability amplitude of the light curves, following the prescriptions in \cite{vaughan2003}, where $x$, $\sigma_{err}$ and $N$ are the count rate, its error and the number of points in the light curve, respectively, and $S^{2}$ is the variance of the light curve. A source is considered to be variable when $\sigma^{2}_{NXS}$ $>0$, i.e. when the intrinsic amplitude of the variability is greater than zero. 
We also report F$_{var}$, i.e., the intrinsic variability amplitude as in  \cite{vaughan2003}.

\begin{figure*}
\centering
\includegraphics[width=0.7\linewidth]{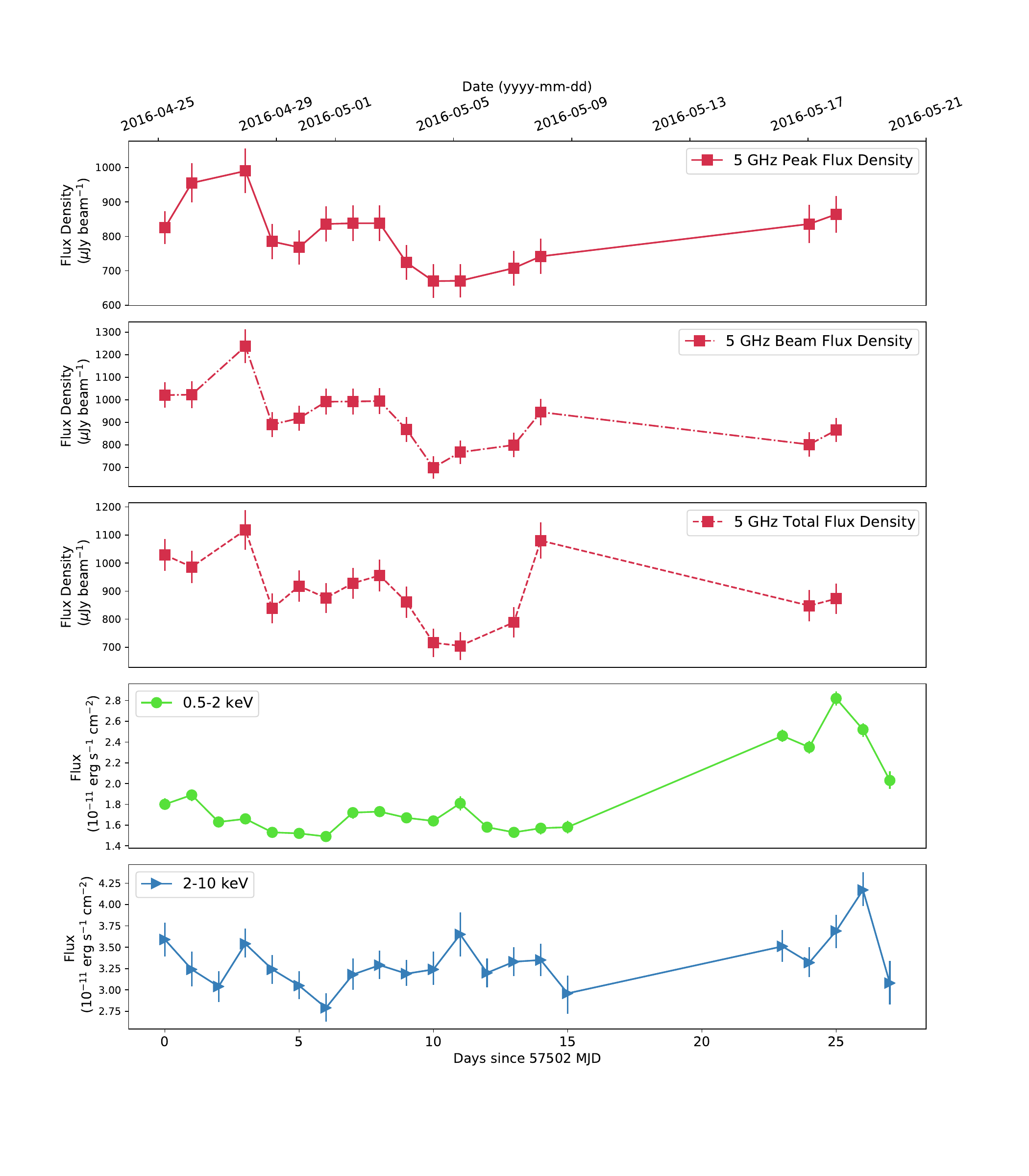}
\caption{Variability is observed in Mrk~110 light curves: 5.0 GHz VLBA peak flux density (top panel), 5.0 GHz VLBA peak flux density, obtained by using a common beam (second panel from top), 5.0 GHz VLBA total flux density (middle panel) and 0.5--2 keV and 2--10 keV X-ray flux (bottom panels, green and blue respectively).}
\label{fig:flux}
\end{figure*}

We tested whether the different synthesized beams at C-band could be responsible for the observed flux density variability in 2016 (where we had daily VLBA observations of Mrk\,110). To this end, we convolved our calibrated data with a common beam whose major and minor axes correspond to the highest values reported in Table~\ref{tab:radio} (i.e., 5.40 mas $\times$ 1.65 mas, with a position angle of zero degrees). As can be seen from Fig.~\ref{fig:flux}, the behaviour and the value of the peak flux densities obtained with the common beam (S$_{beam}$, column 8 in Table~\ref{tab:radio}, second top panel in Fig.~\ref{fig:flux}) are closer to the behaviour of the total flux density (third panel from top) than to those of the peak flux densities obtained with the natural beam (first top panel). This well agrees with a scenario where most, or essentially all, of the flux density comes from the unresolved compact source (less than about 5 mas $\times$ 2 mas in size). This is also reflected in the fact that the variability probability in radio is very similar for the common beam and total flux density, in comparison with the probability for the peak flux density values.

The results of the radio and X-ray light curves of Mrk\,110 are indicative of variations at all frequencies (see Fig.~\ref{fig:flux}). The significance of the radio variability varies from 2.6 $\sigma$ for the peak flux density, to the 3.5 $\sigma$ for the common beam peak flux density and to 3.8 $\sigma$ for the total flux density. The soft X-ray light curve has the maximum variation observed (16 $\sigma$), whereas in the hard X-rays the source shows marginal evidence for variability (2.6 $\sigma$). 
The minimum (maximum) flux of the peak flux density is 670 (990) $\mu$Jy, corresponding to an amplitude variation of 32\% in seven days, while for the total flux density the minimum (maximum) flux is 705 (1118) $\mu$Jy, corresponding to a 37\% variation in eight days. On one day time scale, the most significant ($\sim$ 4 $\sigma$) variability is found between April 27th and 28th, where the total flux density decreases by a factor of 25\%. Significant variability ($\sim$ 21\%) is also observed between May 7th and 8th.

In the 0.5--2 keV energy band, the minimum (maximum) fluxes are 1.5 (2.8)$\times 10^{-11}$ erg s$^{-1}$ cm$^{-2}$, resulting in an increment of 47\% in 19 days, whereas in the 2--10 keV energy band these values were 2.8 (4.2)$\times$ 10$^{-11}$ erg s$^{-1}$ cm$^{-2}$, i.e, an increment of 33\% in 20 days. 

We have estimated the cross-correlation function (CCF) between the radio and the X-ray time series using two methods: Discrete Correlation Function \citep[DCF,][]{dcf1988} and Interpolated Cross Correlation Function \citep[ICCF,][]{iccf1986}, however the low statistics involved prevent us from drawing any conclusion. 

\begin{table*}
\begin{center}
\caption{\bf Variability analysis of the light curves.}

\label{tab:variability}
\begin{tabular}{ l | c c c l c l}
\hline \hline 

  &  Mean & Stddev & $\chi^2$/d.o.f &  1-P$_{var}$ & $\sigma^2_{NXS}$ & F$_{var}$ (\%) \\ \hline
Peak Radio 5 GHz &  803 & 94 & 41/14 &   1.5$\times 10^{-4}$ & 0.0093$\pm$0.0036 & 10$\pm$2\\  
Total Radio 5 GHz &  901 & 120 & 62/14 &   4.5$\times 10^{-9}$ & 0.0139$\pm$0.0040 &12$\pm$2\\  
Common beam Radio 5 GHz &  921 & 132 & 69/14 &   2.7$\times 10^{-9}$ & 0.0168$\pm$0.0044 &13$\pm$2\\  
Calibrator Radio 5 GHz &  198 & 12 & 23/14 &   0.05 & 0.0013$\pm$0.0013 & 4$\pm$1 \\  
X-rays 0.5-2 keV & 1.8$\times 10^{-11}$ & 0.4$\times 10^{-11}$ & 890.1/20 &    0 & 0.0423$\pm$0.0026 & 21$\pm$1 \\
X-rays 2-10 keV & 3.3$\times 10^{-11}$ & 0.3$\times 10^{-11}$ & 51.8/20  & 1.2$\times 10^{-4}$ & 0.0051$\pm$0.0020 & 7$\pm$1\\
 \hline 

\end{tabular}
\end{center}
Table Notes: For each light curve we report the mean flux value and standard deviation (in $\mu$Jy for radio, mJy for the calibrator and erg \hspace{0.1mm} s$^{-1}$cm$^{-2}$ for the X-rays), the $\chi^2$ test with the degrees of freedom, the P$_{var}$ parameter, the normalised excess variance, $\sigma^2_{NXS}$ and the amplitude of variability, F$_{var}$.
\end{table*}

\section{Discussion and conclusions}

We report on daily to weekly radio variability on mas scales in the NLSy1 Mrk~110 via a VLBA monitoring campaign, covering the period from 14 August 2015 till 19 May 2016, with a nearly daily cadence between 24 April and 19 May 2016 (see Table~\ref{tab:radio} and Fig.~\ref{fig:flux}). Simultaneous soft X-ray variability is detected with {\it Swift}/XRT.
This is the first time that radio variability at milli-arcsecond spatial resolution and on a temporal scale of days is tested in an extremely radio faint NLSy1, accreting at 
ten per cent of the Eddington limit (see below). The source shows the most significant variations in the radio 5 GHz total flux density (3.8 $\sigma$) and in the soft X-ray (16 $\sigma$) light curves.

The unresolved radio core in Mrk\,110 shows radio properties consistent with those of other RQ NSLy1 observed with VLBI \citep{doi13}. In particular our results are in agreement with those obtained by \citet{doi13} at 1.4 GHz, both in flux and brightness temperature, suggesting moderate variability on years time scales at mas scales.

The black hole mass is 4.8 $\pm$ 2.0 $\times$ 10$^{7}$ M$_{\odot}$, from a virial mass estimate in \citet{bischoff99}, updated using the virial form factor 1.12 \citep{woo15}.
The X-ray Eddington ratio (assuming an averaged 2--10 keV flux as in Table~\ref{tab:variability}) is L$_{\rm 2-10 keV}$/L$_{\rm Edd}$ $=$ 0.016, (Log L$_{\rm bol}$/L$_{\rm Edd}$ $\sim$ 0.1, assuming the bolometric correction from \citet{duras20}). 
Using the averaged VLBA flux density at 5 GHz and the averaged 2--10 keV luminosity, we have calculated the X-ray radio--loudness parameter \citep{terashima03} as Log L$^{\rm VLBA}_{\rm 5 GHz}$/L$_{\rm 2-10 keV}$ = -5.9. These values place Mrk~110 among the most RQ highly accreting sources in the local Universe (see Fig. 2 \& 3 in \citet{panessa13}).

The high brightness temperature (in the range between 7.5 and 15.7 $\times$ 10$^{7}$ K) allows us to exclude a star-formation origin for the radio emission at the observed spatial scales. Indeed, T$_{B}$ typically lower than 10$^{6}$ K is expected from star-forming regions (e.g., \citealt{pereztorres21}). The emission from supernova remnants and massive stars is averaged on long time scales, therefore not strongly variable \citep{condon13}.
 
From our data, on average, $\sim$ 10\% of the radio emission is variable on a time scale of days.
By applying a simple causality argument, the minimum variability on a time scale of one day implies that such variable 5 GHz radio emission comes from a region of a size smaller than 8.4 $\times$ 10$^{-4}$ pc, i.e., $\sim$ 180 Schwarzschild radii. 
This region likely coincides with the innermost region of the AGN (as suggested by the variability of the peak flux density), but in principle
it could arise from a close-by region, anywhere within $\sim$ 9 pc (as estimated by the radio maximum beam size). 

Assuming that the emitting region is nuclear, the radio variability might originate from the inner accretion disc/corona system or from the base of a low-power jet. The region size is consistent with estimates of time lags from the {\it XMM--Newton} light curve \citep{dasgupta06}, which suggest a size of the X-ray corona of 10--20 R$_{\rm S}$ (typical of RQ AGN, e.g., \citet{kara16}, see however \citet{mastroserio20}). This was also confirmed by recent X-ray/optical lag measurements \citep{vincentelli21, porquet21}.

Both the radio and the X-ray emission may be produced in magnetized plasma in the accretion-disc corona, in analogy with coronally active stars. According to this scenario a {\bf L$_{\rm 5 GHz}$/L$_{\rm X-ray}$} ratio of 10$^{-5}$ is predicted \citep{laor08}. Our ratio -5.9 (in logarithm) falls within the scatter of such relation. In the above scenario we would expect a flat spectral slope, due to synchrotron self-absorption at our frequencies \citep{raginski16}. 
Possible Coronal Mass Ejections (CMEs) in AGN coronae would emit extended optically thin radio emission from outflowing blobs of highly magnetized plasma. In high-Eddington accretion discs, as in our object, an intense coronal activity is indeed expected \citep{laor08,liu14,inoue18}. 
The base of the jet and the corona may be powered by a common magnetic energy reservoir \citep{malzac04}, or in case of highly sub-luminal CMEs, the jet base would physically coincide with the corona \citep{liu14}.
Indeed, at high Eddington ratios, the corona could be outflowing, being slightly collimated near the SMBH and then spreading at larger scales in the form of a diffuse radio emitting plasma \citep{markoff05, king11}.

The estimated spatial scales are also consistent with the BLR scales, as estimated in \citet{liu17, kollatschny03}. However, the free-free contribution of BLR dustless gas at a few GHz is likely negligible, as free-free emission from photo-ionized gas in RQ AGN is expected to dominate the mm range \citep{baskinlaor21}. 

Alternatively, the variable emission may arise from close-by components such as a sub-pc/pc scale jet or outflow. 
For instance, a scenario with an AGN-driven nuclear wind shocking the galaxy medium foresees steep spectral optically thin radio emission \citep{zakamska14}. However, this has been probed mainly at the NLR scales ($>$ 100 pc). 
The presence of variable outflow components very close to the inner region is expected in sources accreting at high Eddington ratios, as observed in X-rays \citep{mizumoto21}. Indeed, the steep spectrum between$\sim$5-8 GHz combined with the high Eddington ratio favour a radiation pressure driven wind scenario \citep{laor19}. However, it is currently not clear how much emission we could expect from nuclear outflowing gas in the radio regime.

The observed brightness temperature is a strong evidence for non-thermal processes acting either in the core or in a nuclear jet knot \citep{ulvestad05}. The jet origin is supported by the observed steepening of spectrum above 5 GHz, ascribable to electronic losses of optically thin synchrotron emission. The lack of mas extended emission suggests a low power sub-relativistic jet confined at $<$ 9 pc scales, which dissipates before leaving the core. The marginal evidence for resolved nuclear emission in the 8th May epoch might support the hypothesis of the presence of a non-nuclear component subject to a brightening episode. However, the quality of our data does not allow to draw any conclusion in this respect.

The variable spectral index (from $\alpha$ $\sim$ -0.5 to -0.15), indicates that the emitting source may transition from an optically thick to an optically thin regime. In the case of an optically thick synchrotron source, according to equation 19 in \citet{laor08}, the resulting emitting region size would be of $\sim$ 0.001 pc, consistent with the region derived from our variability analysis.

We note that the VLBA flux density recovers only 30\% of the VLA-A flux density \footnote{This value has been derived from an averaged VLA 8.4 GHz flux of 1.8 mJy and assuming a 0.7 spectral index to extrapolate to 5 GHz.}, suggesting that some radio emission extends to larger scales, as typically found in RQ Seyferts \citep{panessa13}. Indeed, a secondary component at 3.0$\arcsec$ (2.1 kpc) to the north of a central component was seen in the VLA A-array images \citep{kukula98}, indicating that possibly some past ejection event was able to propagate into the medium without dissipating. The lack of jet imaged at mas scale may be due to observational limitations (e.g., lack sensitivity, uv-coverage). Higher angular resolution (sub-mas scales) and more sensitive (a few $\mu$Jy rms) VLBI images are needed to investigate possible low surface brightness jet-like features, such as those that can be achieved by the Global VLBI \footnote{\url{https://science.nrao.edu/facilities/vlba/docs/manuals/propvlba/array_setups/global-vlbi}}.

A detailed long term JVLA/Swift monitoring programme has been carried out (McHardy et al. in preparation). The angular scales mapped by the JVLA (sub arcsecond) will be larger with respect to those of VLBI (milli-arcsecond), however, the time scale of a few months will allow a significant cross correlation analysis, shedding light to the origin of radio emission and of the accretion and ejection coupling in this NLSy1. As an example, the detection of an X-ray/radio correlation with a $\sim$ tens of days radio lag is expected if the radio emission comes from a synchrotron jet, such as that seen in low accretion rate X-ray binary systems (e.g. \citet{corbel13}). We refer to \citet{panessa19} for a more extensive discussion.

Currently, radio variability in RQ AGN is largely unexplored. We encourage multi-frequency monitoring campaigns of RQ AGN to include radio light curves, as the strong coupling between the accretion and the ejection phenomena is the key for their comprehension.

\section*{Acknowledgements} 

FP acknowledges support from a grant PRIN-INAF SKA-CTA 2016 and from Agenzia Spaziale Italiana (ASI) under contract n. 2019-35-HH.0. 
MPT acknowledges financial support from the State Agency for Research of the Spanish MCIU through the ``Center of Excellence Severo Ochoa'' award to the Instituto de Astrof\'isica de Andaluc\'ia (SEV-2017-0709)
and through grants PGC2018-098915-B-C21 and PID2020-117404GB-C21 (MCI/AEI/FEDER, UE).
LHG acknowledges funds by ANID – Millennium Science Initiative Program – ICN12$\_$009 awarded to the Millennium Institute of Astrophysics (MAS). 
 
\bibliographystyle{aasjournal}
\bibliography{mrk110}{}

\section*{Data Availability} 

The data underlying this article were accessed from NRAO and Swift (https://science.nrao.edu/ and https://swift.gsfc.nasa.gov/). The derived data generated in this research will be shared on reasonable request to the corresponding author.

 \end{document}